\documentclass[ALICE,manyauthors]{cernphprep}
\usepackage[comma,square,numbers,sort&compress]{natbib}
\usepackage{hyperref}
\usepackage{lineno}
\usepackage{xspace}
\usepackage{amssymb}
\usepackage{amsmath,bm}
\usepackage{mathrsfs}
\usepackage{epstopdf}
\usepackage[T1]{fontenc}
\usepackage{orcidlink}

\newcommand{\PbPb}         {\mbox{Pb--Pb}\xspace}

\newcommand{\pPb}          {\mbox{p--Pb}\xspace}

\newcommand{\pt}           {\ensuremath{p_{\rm T}}\xspace}

\newcommand{\RpPb}         {\ensuremath{R_{\rm pPb}}\xspace}

\newcommand{\nineH}        {$\sqrt{s}~=~0.9$~Te\kern-.1emV\xspace}
\newcommand{\seven}        {$\sqrt{s}~=~7$~Te\kern-.1emV\xspace}
\newcommand{\twoH}         {$\sqrt{s}~=~0.2$~Te\kern-.1emV\xspace}
\newcommand{\twosevensix}  {$\sqrt{s}~=~2.76$~Te\kern-.1emV\xspace}
\newcommand{\five}         {$\sqrt{s}~=~5.02$~Te\kern-.1emV\xspace}
\newcommand{\twosevensixnn}{$\sqrt{s_{\mathrm{NN}}}~=~2.76$~Te\kern-.1emV\xspace}
\newcommand{\fivenn}       {$\sqrt{s_{\mathrm{NN}}}~=~5.02$~Te\kern-.1emV\xspace}

\newcommand{\GeV}          {\text{Ge\kern-.1emV}\xspace}
\newcommand{\MeV}          {\text{Me\kern-.1emV}\xspace}
\newcommand {\tev}      {\text{Te\kern-.1emV}\xspace}
\newcommand{\TeV} {\tev}

\newcommand{\GeVc}         {\ensuremath{\GeV/c}\xspace}

\newcommand{\Lint}         {\ensuremath{\mathcal{L}_\mathrm{int}}\xspace}

\newcommand{\ee}           {\ensuremath{\mathrm{e^{+}e^{-}}}\xspace} 
\newcommand{\epem}{\ee}

\newcommand {\pT}        {\pt}
\newcommand {\meanpT}    {\ensuremath{\langle p_{\mathrm{T}} \kern-0.1em\rangle}\xspace}
\newcommand {\mean}[1]   {\ensuremath{\langle #1 \kern-0.1em\rangle}\xspace} 
\newcommand {\sqrtsNN}   {\ensuremath{\sqrt{s_{\textsc{NN}}}}\xspace}

\newcommand {\ep}        {\mbox{$\mathrm {e^-p}$}\xspace}

\newcommand {\MeanNpart} {\mbox{\ensuremath{< \kern-0.15em N_{part} \kern-0.15em >}}}

\newcommand {\mmom}     {\mbox{\rm MeV$\kern-0.15em /\kern-0.12em c$}}
\newcommand {\gmom}     {\mbox{\rm GeV$\kern-0.15em /\kern-0.12em c$}}
\newcommand {\mmass}    {\mbox{\rm MeV$\kern-0.15em /\kern-0.12em c^2$}}

\newcommand {\dg}       {\mbox{$\kern+0.1em ^\circ$}}

\newcommand{\mub}{\ensuremath{\mu\rm b}\xspace}
\newcommand{\mb}{\ensuremath{\rm mb}\xspace}

\newcommand{\Lb}{\ensuremath{\rm {\Lambda_b^{0}}}\xspace}
\newcommand{\Xic}         {\ensuremath{\mathrm {\Xi_{c}}}\xspace}

\newcommand{\lambdac}     {\ensuremath{\mathrm {\Lambda_{c}^{+}}}\xspace}
\newcommand{\xicz}        {\ensuremath{\mathrm {\Xi_{c}^{0}}}\xspace}
\newcommand{\xiczp}        {\ensuremath{\mathrm {\Xi_{c}^{0,+}}}\xspace}
\newcommand{\Xicz}{\xicz}
\newcommand{\Xiczp}{\xiczp}
\newcommand{\XicD} {\ensuremath{\xiczp/\Dz}\xspace}
\newcommand{\xicp}        {\ensuremath{\mathrm {\Xi_{c}^{+}}}\xspace}
\newcommand{\Xicp}{\xicp}

\newcommand{\rmLambdas}         {\ensuremath{\mathrm {\Lambda \kern-0.2em + \kern-0.2em \overline{\Lambda}}}\xspace}

\newcommand{\Kzs}               {\ensuremath{\mathrm {K^0_S}}\xspace}

\newcommand{\Jpsi}              {\ensuremath{\rm J/\psi}\xspace}

\newcommand{\Dzero}{\ensuremath{\mathrm {D^0}}\xspace}
\newcommand{\Dz}{\Dzero}

\newcommand{\Dstar}{\ensuremath{\rm D^{*+}}\xspace}
\newcommand{\Dplus}{\ensuremath{\rm D^+}\xspace}
\newcommand{\Dp}{\Dplus}
\newcommand{\Dsubs}{\ensuremath{\rm D_{s}^+}\xspace}
\newcommand{\Ds}{\Dsubs}

\newcommand{\Lcplus}{\lambdac}
\newcommand{\Lc}         {\Lcplus}
\newcommand{\LcD} {\ensuremath{\lambdac/\Dzero}\xspace}

\newcommand{\LctopKpi}{\ensuremath{\rm \Lambda_{c}^{+}\to p K^-\pi^+}\xspace}

\newcommand{\LctopKzS}{\ensuremath{\rm \Lambda_{c}^{+}\to p K^{0}_{S}}\xspace}

\newcommand{\LctopKs}{\LctopKzS}

\newcommand{\sqrtsfive}{\ensuremath{\sqrt{s} = 5.02~\TeV}\xspace}
\newcommand{\sqrtsthirt}{\ensuremath{\sqrt{s} = 13~\TeV}\xspace}

\newcommand{\sqrtsNNfive}{\ensuremath{\sqrt{s_\mathrm{NN}} = 5.02~\TeV}\xspace}

\newcommand{\figref}[1]{Fig.~\ref{#1}}

\newcommand{\tabref}[1]{Table~\ref{#1}}

\newcommand{\Omegac}{\ensuremath{\Omega_{\rm c}^{0}}\xspace}

\newcommand{\ccbar}{\ensuremath{\rm c\overline{c}}\xspace}

\newcommand{\hc}{\ensuremath{\rm h_{c}}\xspace}
\newcommand{\fragfraction}{\ensuremath{f(\mathrm{c}\rightarrow\hc)}\xspace}
\newcommand{\dsigccdy}{\ensuremath{\mathrm{d}\sigma({\ccbar})/\mathrm{d}y}\xspace}

\newcommand{\dsigmadpt}{\ensuremath{\mathrm{d}\sigma/\mathrm{d}\pt}\xspace}
\newcommand{\dsigmady}{\ensuremath{\mathrm{d}\sigma/\mathrm{d}y}\xspace}

\newcommand{\ValUncStatSystUD}[4]{\ensuremath{#1 \pm #2\,\mathrm{(stat.)} {}^{+#3}_{-#4}\,\mathrm{(syst.)}}\xspace}

\begin{document}

\begin{titlepage}
\PHyear{2024}       
\PHnumber{141}      
\PHdate{22 May}  

\title{Charm fragmentation fractions and $\mathbf{c\overline{c}}$ cross section in p--Pb collisions at $\mathbf{\sqrt{\emph{s}_{NN}}=5.02\;TeV}$}
\ShortTitle{Charm cross section and fragmentation fractions in p--Pb collisions}   

\Collaboration{ALICE Collaboration\thanks{See Appendix~\ref{app:collab} for the list of collaboration members}}
\ShortAuthor{ALICE Collaboration} 

\begin{abstract}
The total charm-quark production cross section per unit of rapidity \dsigccdy, and the fragmentation fractions of charm quarks to different charm-hadron species \fragfraction, are measured for the first time in \pPb collisions at \sqrtsNNfive at midrapidity ($-0.96<y<0.04$ in the centre-of-mass frame) using data collected by ALICE at the CERN LHC. The results are obtained based on all the available measurements of prompt production of ground-state charm-hadron species: $\mathrm{D}^{0}$, $\mathrm{D}^{+}$, $\mathrm{D}_\mathrm{s}^{+}$, and $\mathrm{J/\psi}$ mesons, and $\Lambda_\mathrm{c}^{+}$ and $\Xi_{\rm c}^{0}$
baryons. The resulting cross section is 
$ \dsigccdy =219.6 \pm 6.3\;(\mathrm{stat.}) {\;}_{-11.8}^{+10.5}\;(\mathrm{syst.}) {\;}_{-2.9}^{+8.3}\;(\mathrm{extr.})\pm 5.4\;(\mathrm{BR})\pm 4.6\;(\mathrm{lumi.}) \pm 19.5\;(\text{rapidity shape})+15.0\;(\Omegac)\;\mb$, 
which is consistent with a binary scaling of pQCD calculations from pp collisions. The measured fragmentation fractions are compatible with those measured in pp collisions at $\sqrt{s} = 5.02$ and $13$\;TeV, showing an increase in the relative production rates of charm baryons with respect to charm mesons in pp and p--Pb collisions compared with $\mathrm{e^{+}e^{-}}$ and $\mathrm{e^{-}p}$ collisions. The $p_\mathrm{T}$-integrated nuclear modification factor of charm quarks, $R_\mathrm{pPb}(\ccbar)=  0.91 \pm 0.04\;{\rm (stat.)}{}^{+0.08}_{-0.09}\;{\rm (syst.)}{}^{+0.05}_{-0.03}\;{\rm (extr.)}{}\pm 0.03\;{\rm (lumi.)}$, is found to be consistent with unity and with theoretical predictions including nuclear modifications of the parton distribution functions. 

\end{abstract}
\end{titlepage}

\setcounter{page}{2} 


\section{Introduction}  \label{sec:intro}

The processes governing the production of heavy-flavour hadrons (those containing at least one charm or beauty quark) in hadronic collisions have recently become a focal point of study at the CERN LHC. Due to their large masses ($m_\mathrm{c,b} \gg \Lambda_\mathrm{QCD}$), heavy-flavour hadron production is calculable using perturbative quantum chromodynamics (pQCD) under the assumption of factorisation~\cite{Collins:1989gx,Catani:1990eg}. In this approach, the production cross section of a given hadron species is considered as a convolution of: (i) the parton distribution functions (PDFs) of the colliding hadrons; (ii) the partonic hard-scattering cross sections for a heavy quark to be produced; and (iii) the fragmentation functions, which describe the fraction of the heavy-quark momentum carried by the heavy-flavour hadron. The fragmentation term in the calculation also governs the relative abundances of each hadron species, which are often referred to as `fragmentation fractions'. 
Typically, in the factorisation approach, the hadronisation process is taken to be independent of the collision system. Therefore, the fragmentation functions measured in \ee collisions are applied equally to calculations in proton--proton (pp) collisions. The assumption of fragmentation as a universal hadronisation mechanism would also result in identical fragmentation fractions between \epem and pp collisions. This has been tested extensively by measuring the production cross section ratios of different charm-hadron species.

Measurements of the $\Dp/\Dz$ and $\Ds/\Dz$ meson-to-meson production ratios in pp collisions at different energies by the ALICE Collaboration~\cite{ALICE:2021dhb,ALICE:charmfrag13} have shown consistent production rates of charm mesons between hadronic collisions and earlier measurements from \epem and \ep collisions~\cite{Lisovyi:2015uqa}. However, more recent measurements of the \LcD and \XicD production ratios in pp and heavy-ion collisions~\cite{ALICE:charmfrag13,ALICE:2022exq,ALICE:XicpPb,LHCb:2018weo,LHCb:2023cwu,LHCb:2022ddg,CMS:2019uws} have shown a marked increase in charm-baryon production with respect to charm-meson production over the \epem baseline in hadronic collisions at LHC energies. This was further quantified by determining the charm-hadron fragmentation fractions in pp collisions at \sqrtsfive~\cite{ALICE:2021dhb,ALICE:charmfrag13} and \sqrtsthirt~\cite{ALICE:charmfrag13}. In both cases, a significant relative enhancement of charm-baryon production was observed, along with a corresponding depletion in the charm-meson fragmentation fractions. The results are also consistent within uncertainties between \sqrtsfive and \sqrtsthirt, implying that the collision energy does not play a significant role in the relative abundances of charm-hadron species in hadronic collisions at the LHC. The increase in charm-baryon production is also reflected in an increase of the total charm production cross section in pp collisions over earlier determinations using only the measured D-meson cross sections with the fragmentation fractions from leptonic colliders~\cite{PhysRevC.84.044905,PhysRevD.86.072013,PhysRevLett.94.062301}. More recently, efforts have also been made to extrapolate and combine the available measurements of charm hadrons in pp collisions from different LHC experiments in complementary regions of phase space in order to determine the total charm production cross section~\cite{Bierlich:2023ewv}.

Measurements of the charm fragmentation fractions in \pPb collisions are also of particular interest, as they reveal whether the observed changes in relative hadronisation rates for \epem and pp collisions are mainly due to the increase of the particle multiplicity between these collision systems. The mean charged-particle multiplicity per unit of pseudorapidity $\left.\left\langle\mathrm{d}N_\mathrm{ch}/\mathrm{d}\eta\right\rangle\right|_{|\eta| < 1.0}$ in collisions at the LHC increases both as a function of energy and system size -- for minimum-bias pp collisions at \sqrtsfive,  $\left.\left\langle\mathrm{d}N_\mathrm{ch}/\mathrm{d}\eta\right\rangle\right|_{|\eta| < 1.0} = 5.48\pm0.05\;(\text{uncorr.\,syst.})\pm0.05\;(\text{corr.\,syst.})$~\cite{ALICE:2020swj}, and for minimum-bias \pPb collisions at \sqrtsNNfive, $\left.\left\langle\mathrm{d}N_\mathrm{ch}/\mathrm{d}\eta\right\rangle\right|_{|\eta| < 1.0}=16.81\pm0.71\;({\rm syst.})$~\cite{ALICE:2012xs}. 
The ALICE Collaboration measured the \LcD yield ratio in intervals of charged-particle multiplicity in pp collisions at \sqrtsthirt~\cite{ALICE:2021npz}, showing that in a hadronic collision, even at relatively small multiplicities, charm-quark hadronisation proceeds differently from \ee collisions. In the beauty sector, the measurements of \Lb-baryon production relative to that of B mesons at forward rapidity in pp and \pPb collisions by the LHCb Collaboration show a modification of the fragmentation fractions among collision systems similar to that observed for charm quarks at midrapidity~\cite{LHCb:2019fns,LHCb:2015qvk, LHCb:2023wbo, LHCb:2019avm}. The ALICE Collaboration measured production cross sections of \Dzero and \Lc hadrons originating from beauty-hadron decays at midrapidity in proton--proton collisions at a centre-of-mass energy $\sqrt{s}$ = 13 TeV~\cite{ALICE:2023wbx}, and a baryon-to-meson enhancement similar to the one measured by LHCb at forward rapidity was observed.

The study of charm production in \pPb collisions serves as an additional baseline to examine so-called cold nuclear matter (CNM) effects at the LHC. Due to different effects related to the presence of a nucleus in the colliding system, there may also be corresponding changes to the yield and transverse momentum (\pt) distribution of charm hadrons. These effects include modifications of the PDFs in nuclei (nPDFs) with respect to the proton PDFs~\cite{Arneodo:1992wf,Klasen:2023uqj} or $k_{\rm T}$-broadening due to multiple soft collisions before the heavy-quark pair is produced~\cite{Lev:1983hh,Kopeliovich:2002yh}.  The resulting modifications of the yields are typically examined using the nuclear modification factor \RpPb~\cite{ALICE:2019fhe}. The ALICE measurements of \Lc production in pp and \pPb collisions at \sqrtsNNfive~\cite{ALICE:2022exq} revealed a significant modification of the \pT distribution of the charm baryon-to-meson ratio between \pPb and pp collisions, which was not observed for the charm meson-to-meson ratios within the current measurement uncertainties. In addition, an increase of the mean transverse momentum of \Lc baryons was observed in \pPb collisions with respect to pp collisions, with no significant difference seen for \Dz mesons. This hardening of the \pT spectrum of \Lc baryons in \pPb collisions indicates a possible influence of radial flow~\cite{Bozek:2011if,Weller:2017tsr} and/or hadronisation via coalescence~\cite{Song:2018tpv, Minissale:2020bif,Beraudo:2023nlq}. These effects are consistent with observations from the light-flavour sector, such as measurements of the $\Lambda/\Kzs$ baryon-to-meson ratio in \pPb collisions~\cite{ALICE:2013wgn}. 
However, the \pT-integrated \RpPb of \Lc baryons was found to be consistent with that of \Dz mesons and with unity. 
With the recent addition of \Xicz baryons to the available measurements~\cite{ALICE:XicpPb}, it has become possible to study potential nuclear modifications of the total \ccbar production cross section to conclude whether nPDFs affect the overall creation of charm hadrons. A recent review with a more comprehensive overview of heavy-quark hadronisation can be found in Ref.~\cite{Altmann:2024kwx}.

In this article, the first measurements of the total charm production cross section per unit of rapidity and the charm-hadron fragmentation fractions at midrapidity in \pPb collisions at the LHC are presented. Section~\ref{sec:analysis} outlines the previous measurements by the ALICE Collaboration that were used to compute the total charm-quark production cross section and the fragmentation fractions of charm quarks to different charm-hadron species. It also details the theoretical models that were used to extrapolate the measured charm-hadron production cross sections down to $\pt=0$, where necessary. Section~\ref{sec:results} shows the results for the charm-hadron fragmentation fractions and \ccbar production cross section in \pPb collisions at \sqrtsNNfive, along with the \pt-integrated nuclear modification factor \RpPb. A summary is given in Section~\ref{sec:summary}.

\section{Dataset and analysis} \label{sec:analysis}

The ALICE detector system is described in detail in Ref.~\cite{ALICE:2008ngc}. The measured prompt charm-hadron production cross sections used in this article were obtained from a sample of minimum-bias \pPb collisions collected in 2016 during Run 2 of the LHC with a total integrated luminosity of $\Lint=287\pm11\;\mub^{-1}$~\cite{ALICE:2014gvw}. The measurements were performed at midrapidity (defined as $-0.96<y<0.04$ in the collision centre-of-mass frame).  Prompt charm hadrons are those produced directly in the hadronisation of a charm quark or via the decay of a directly produced excited charm-hadron or charmonium state, while charm hadrons produced by the decays of beauty hadrons are denoted as `non-prompt'. D mesons were reconstructed in the decay channels $\Dz\to\mathrm{K}^-\pi^+$, $\Dp\to\mathrm{K}^-\pi^+\pi^+$, $\Ds\to\phi\pi^+$, and $\Dstar\to\Dz\pi^+$ (and respective charge conjugates)~\cite{ALICE:2019fhe}. Charm baryons were reconstructed in the channels \LctopKs~\cite{ALICE:2022exq,ALICE:2020wfu}, \LctopKpi~\cite{ALICE:2020wfu}, and $\Xicz\to\Xi^-\pi^+$~\cite{ALICE:XicpPb} (and respective charge conjugates). The \Jpsi mesons were reconstructed in the channel $\Jpsi\to\epem$~\cite{ALICE:2021lmn}. The details of the hadron reconstruction methods can be found in the corresponding citations.

In order to derive the total charm production cross section and the individual charm-hadron fragmentation fractions at midrapidity, the \pt-integrated rapidity-differential cross sections must be considered. For \Dz mesons~\cite{ALICE:2019fhe} and \Lc baryons~\cite{ALICE:2022exq}, the \pt-differential yields were measured down to $\pt=0$ and the calculation of the \pt-integrated cross section did not require further extrapolation. For the remaining hadron species, where no measurement down to $\pt=0$ is available, an extrapolation was performed based on model calculations.

For \Dplus, \Ds, and \Dstar mesons, POWHEG+PYTHIA~6 calculations were used to extrapolate the \pt spectra, which were measured down to $\pt=1$\;\GeVc for \Dplus and \Dstar, and $\pt=2$\;\GeVc for \Ds. 
POWHEG is an event generator with next-to-leading order (NLO) accuracy~\cite{Frixione:2007nw}. It is matched with PYTHIA~6~\cite{Sjostrand:2006za} to generate the parton shower and fragmentation. The CT14NLO parton distribution functions~\cite{Dulat:2015mca} with the EPPS16 parameterisation of the nPDFs~\cite{Eskola:2016oht} were used in the calculations. The POWHEG+PYTHIA~6 simulations describe the charm-meson production cross sections within uncertainties. 
The extrapolation factors were determined from the fraction of the theoretically calculated cross section in the visible region with respect to the \pt-integrated one, relying on the \pt shape of the production spectra without directly using the magnitude from theory. This was performed by taking the integral of the calculated production cross section for $0<\pt<50\;\GeVc$ and dividing it by the corresponding integral in the measured \pt region for that species. The contribution to the total cross section for $\pt>50\;\GeVc$ is negligible with respect to that for $\pt<50\;\GeVc$, so any extrapolation to higher \pT values than this would have no impact on the final result. An uncertainty on this extrapolation factor was then determined by separately performing the same procedure using the following standard variations on the POWHEG+PYTHIA~6 model: (i) varying the factorisation and renormalisation scales in the POWHEG simulation ($\mu_\mathrm{r,f}$) within the range $0.5<\mu_{\mathrm{r,f}}/\mu_0<2.0$ (where $\mu_0 = \sqrt{m^2+\pt^2}$), with the additional constraint $0.5<\mu_\mathrm{r}/\mu_\mathrm{f}<2$; and (ii) considering differences due to the uncertainty sets on the EPPS16 nPDF parametrisations.
The maximum relative upward and downward differences from the central extrapolation factor provided by (i) and (ii) were added in quadrature to give the overall extrapolation uncertainty on the cross sections, which amounted to ${}^{+7.7}_{-2.2}\%$ for \Dplus, ${}^{+18.8}_{-\phantom{0}7.3}\%$ for \Ds, and ${}^{+27.0}_{-\phantom{0}2.3}\%$ for \Dstar mesons. In all cases, the contribution due to the values of $\mu_\mathrm{r,f}$ was much larger than that from the nPDF uncertainty sets. The extrapolation factors were $1.27^{+0.10}_{-0.04}$ for \Dplus mesons, $1.25^{+0.08}_{-0.04}$ for \Dstar mesons, and $2.08^{+0.40}_{-0.20}$ for \Dsubs mesons.

For \Xicz baryons, the extrapolation was performed and reported in Ref.~\cite{ALICE:XicpPb}.
The POWHEG+PYTHIA~6 calculations were found to poorly describe the \pt shape of the measured differential cross sections of charm baryons in \pPb collisions. Instead, for the case of \Xicz, the quark (re)combination model (QCM)~\cite{Li:2017zuj,Li:2021nhq} was chosen to perform the extrapolation, as it is tuned to reproduce previous measurements of \Lc-baryon production from ALICE and provides a reasonable description of the shape of the measured \dsigmadpt of \Xicz baryons~\cite{ALICE:XicpPb}. As the QCM model does not provide any theoretical uncertainties, the extrapolation uncertainty was defined by adding in quadrature (i) the maximum difference between the central extrapolation factor determined using QCM and those calculated using PYTHIA~8 with the Angantyr generator~\cite{Bierlich:2018xfw} and POWHEG+PYTHIA~6 (${}^{+20.6}_{-\phantom{0}0.0}\%$), and (ii) the envelope of the POWHEG+PYTHIA~6 variations for \Xicz as described above for the D mesons (${}^{+23.0}_{-\phantom{0}8.6}\%$). The resulting extrapolation uncertainty is ${}^{+30.9}_{-\phantom{0}8.6}\%$ of the total \Xicz production cross section at midrapidity. The final extrapolation factor for \Xicz baryons is $1.74^{+0.54}_{-0.15}$. 
As no measurement of the \Xicp production cross section is available in \pPb collisions at midrapidity, the contribution of \Xicp to the total charm cross section was considered as equal to that of \Xicz baryons, as also done in Refs.~\cite{ALICE:2021dhb,ALICE:charmfrag13}. This assumption was made based on isospin symmetry, and is further supported by the measurements of \Xiczp in pp collisions at \sqrtsthirt~\cite{ALICE:charmfrag13}, which showed that the production cross sections are consistent between the two charge states.

For \Jpsi mesons, the extrapolation was performed and reported in Ref.~\cite{ALICE:2021lmn}. While the inclusive \Jpsi-meson cross section (defined as the sum of prompt and non-prompt \Jpsi mesons) was measured for ${\pt>0}$, an extrapolation was required for the non-prompt \Jpsi cross section to account for the \pt region ${0<\pt<1\;\GeVc}$ in order to extract the prompt cross section. The extrapolation was performed using Fixed Order plus Next-to-Leading Logarithm (FONLL) pQCD calculations~\cite{fonllcalc1,fonllcalc2}, with the standard CTEQ6.6 PDFs~\cite{Nadolsky:2008zw} modified according to the EPPS16 nPDF parametrisation~\cite{Eskola:2016oht} to describe \pPb collisions. The corresponding value for the prompt component is obtained as the difference between the inclusive
\Jpsi-meson cross section and the non-prompt one, as determined with the extrapolation procedure described above. The \dsigmady value for prompt \Jpsi-meson production from Ref.~\cite{ALICE:2021lmn} was used directly in this work. 

In the calculation of the total charm-hadron yield, the contribution from \Omegac baryons must also be considered. Currently, no measurements of the \Omegac are available in \pPb collisions. In addition, as there is no measurement of the absolute branching ratios for any of the decay modes of the \Omegac baryon, it is not yet possible to directly measure its production cross section and so it is not included in the sum of the charm-hadron cross sections. The potential contribution from \Omegac baryons was instead taken into account by assuming an extreme upper limit of $\sigma(\Omegac) = \sigma(\Xicz)$, and assigning this as a fully asymmetric uncertainty bound on the total charm-hadron yield, as was previously done in Refs.~\cite{ALICE:2021dhb,ALICE:charmfrag13}. The uncertainty related to the estimated \Omegac-baryon production cross section is added in quadrature to the upper extrapolation uncertainty on the total cross section. There are further contributions to the total charm cross section from doubly- and triply-charmed baryons such as $\Xi_\mathrm{cc}^{+}$ and $\Omega_\mathrm{ccc}^{++}$ as well as charmonium states that do not decay to \Jpsi mesons. However, these are considered to be negligible compared to the precision of the current measurement of the total cross section.

When calculating the total \ccbar production cross section, additional correction factors were applied to account for the differing rapidity distributions between charm hadrons, individual charm quarks, and \ccbar pairs. These were computed using the same procedure as in Refs.~\cite{ALICE:charmfrag13, ALICE:2021dhb}. The scaling factor between charm hadrons and single charm quarks was determined from FONLL calculations of pp collisions at \sqrtsfive~\cite{Cacciari:1998it} and found to be unity, with a 2\% uncertainty assigned based on the differences between FONLL and PYTHIA 8 calculations. The scaling between single charm quarks and charm-anticharm pairs was determined using POWHEG+PYTHIA~6 calculations. In the range $-0.96<y<0.04$, this correction factor was found to have a value of 1.03. An uncertainty was assigned by adding in quadrature the envelopes defined by (i) variations of the $\mu_\mathrm{r,f}$ scale parameters in POWHEG and (ii) the uncertainty sets of the EPPS16 nPDFs, with a final value of $8.7\%$. Combining these two factors, the uncertainty on the total charm cross section due to the rapidity shape is 8.9\%.

In calculating the fragmentation fractions and total \ccbar production cross section at midrapidity, the systematic uncertainties related to tracking, feed-down from beauty-hadron decays, \pt extrapolation, and luminosity were propagated as fully correlated among different charm-hadron species; all other uncertainties were propagated as uncorrelated.

\section{Results} \label{sec:results}

The \pt-integrated $\left.\mathrm{d}\sigma(\mathrm{h}_\mathrm{c})/\mathrm{d}y\right|_{-0.96<y<0.04}$ values for all of the considered charm-hadron species in \pPb collisions at \sqrtsNNfive are listed in~\tabref{tab:dsigdys}. 
For all species, the systematic uncertainties related to the extrapolation (extr.), luminosity determination (lumi.), and branching ratio (BR) are reported separately from the systematic uncertainties related to the data analysis, except where these values were combined or not applicable in their original publications.

\begin{table}[h!tb]
    \centering
    \caption{The \pt-integrated rapidity-differential cross sections for all measured charm-hadron species at midrapidity in \pPb collisions at \sqrtsNNfive. For \Lc (\Jpsi) hadrons, the published systematic uncertainties include the contributions from the BR (BR and luminosity) uncertainties.
    }\label{tab:dsigdys}
    \begin{tabular}{c l}
        \multicolumn{1}{c}{$\mathrm{h}_\mathrm{c}$} & \multicolumn{1}{c}{$\left.\mathrm{d}\sigma(\mathrm{h}_\mathrm{c})/\mathrm{d}y\right|_{-0.96<y<0.04} (\mb)$, $\pt>0$} \\\hline\hline
        \rule{0pt}{1.004\normalbaselineskip}\Dz & $88.5\pm2.7\;{\rm (stat.)}{}^{+5.3}_{-6.1}\;{\rm (syst.)}{}\pm 3.3\;{\rm (lumi.){}\pm 0.9\;{\rm (BR)}}$~\cite{ALICE:2019fhe} \\
        \rule{0pt}{1.004\normalbaselineskip}\Dp & $39.1 \pm 2.4\;{\rm (stat.)}{}^{+3.7}_{-3.7}\;{\rm (syst.)}{}^{+3.0}_{-1.1}\;{\rm (extr.)}{}\pm1.4\;{\rm (lumi.)}{}\pm 1.0\;{\rm (BR)}$\\
        \rule{0pt}{1.004\normalbaselineskip}\Ds & $19.2 \pm 1.0\;{\rm (stat.)}{}^{+1.5}_{-1.6}\;{\rm (syst.)}{}^{+3.6}_{-1.8}\;{\rm (extr.)}{}\pm0.7\;{\rm (lumi.)}{}\pm0.7\;{\rm (BR)}$ \\
        \rule{0pt}{1.004\normalbaselineskip}\Lc & $36.9\pm3.3\;{\rm (stat.)}{}\pm 4.5\;{\rm (syst.)}{}\pm 1.4\;{\rm (lumi.)}$~\cite{ALICE:2022exq}\\
        \rule{0pt}{1.004\normalbaselineskip}\Xicz & $15.0 \pm 2.8\;{\rm (stat.)}{}^{+2.2}_{-2.3}\;{\rm (syst.)}{}^{+4.6}_{-1.3}{\rm (extr.)}{}\pm 0.6\;{\rm (lumi.)}{}\pm 3.3\;{\rm (BR)}$~\cite{ALICE:XicpPb}\\
       \rule{0pt}{1.004\normalbaselineskip}\rule[-1.5ex]{0pt}{0pt} \Jpsi & $0.8731\pm 0.0336\;{\rm (stat.)}{}\pm 0.0504\;{\rm (syst.)}{}^{+0.0016}_{-0.0028}\;{\rm (extr.)}$~\cite{ALICE:2021lmn} \\ \hline 
        \rule{0pt}{1.004\normalbaselineskip} \Dstar & $27.7\pm 1.4\;{\rm (stat.)}{}^{+1.4}_{-1.5}\;{\rm (syst.)}{}^{+1.9}_{-0.8}{\rm (extr.)}{}\pm 1.0\;{\rm (lumi.)}{}\pm 1.0\;{\rm (BR)} $
    \end{tabular}
\end{table}

The fragmentation fractions \fragfraction were computed from the integrated production cross sections of the charm hadrons according to

\begin{equation}
\fragfraction=\frac{\mathrm{d}\sigma(\mathrm{h}_\mathrm{c})/{\rm d}{\it y}|_{-0.96<{\it y}<0.04}}{\sum\limits_{\mathrm{h_{c}}}{\mathrm{d}\sigma({\mathrm{h}_\mathrm{c}})/\mathrm{d}{\it y}|_{-0.96<{\it y}<0.04}}}.
\end{equation}

\begin{figure}[h!tb]
    \centering
    \includegraphics[width=0.54\textwidth]{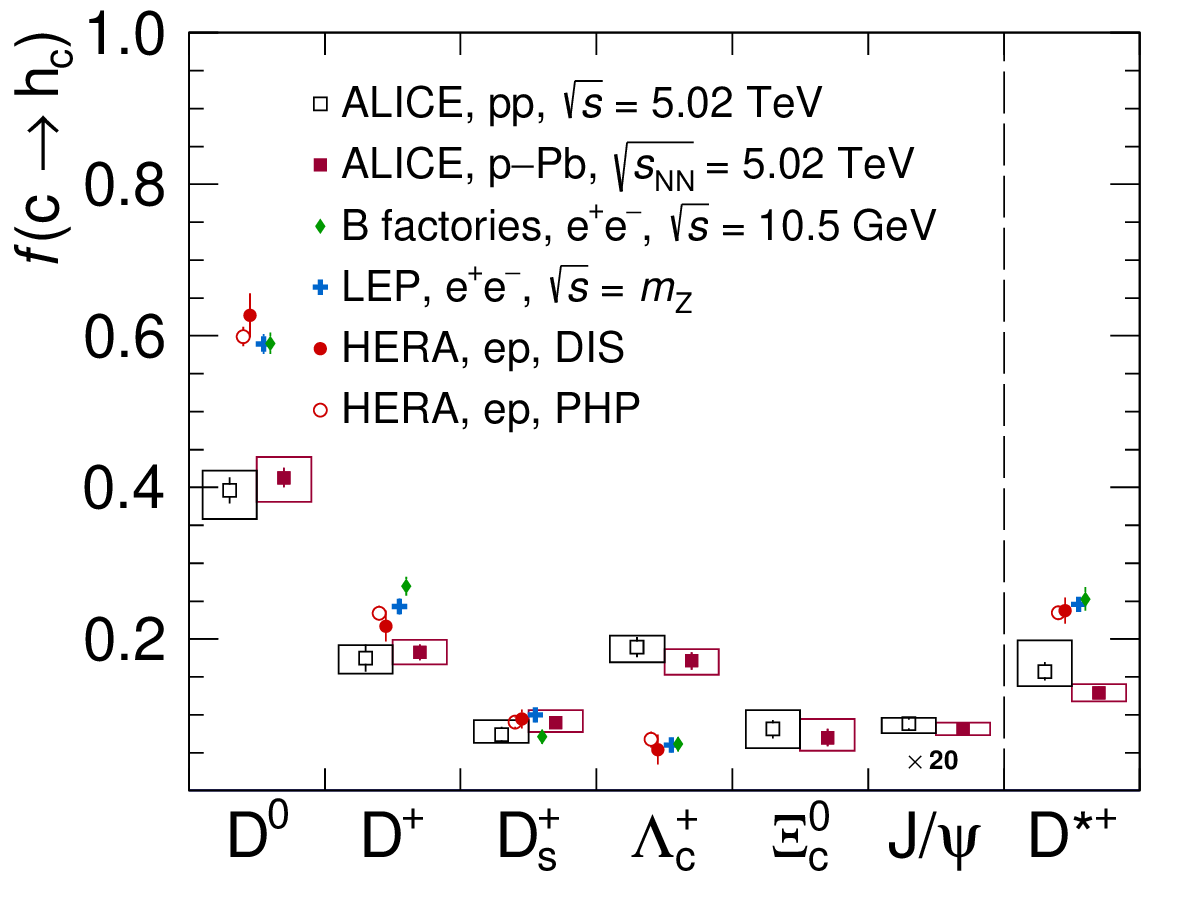}
    \includegraphics[width=0.44\textwidth]{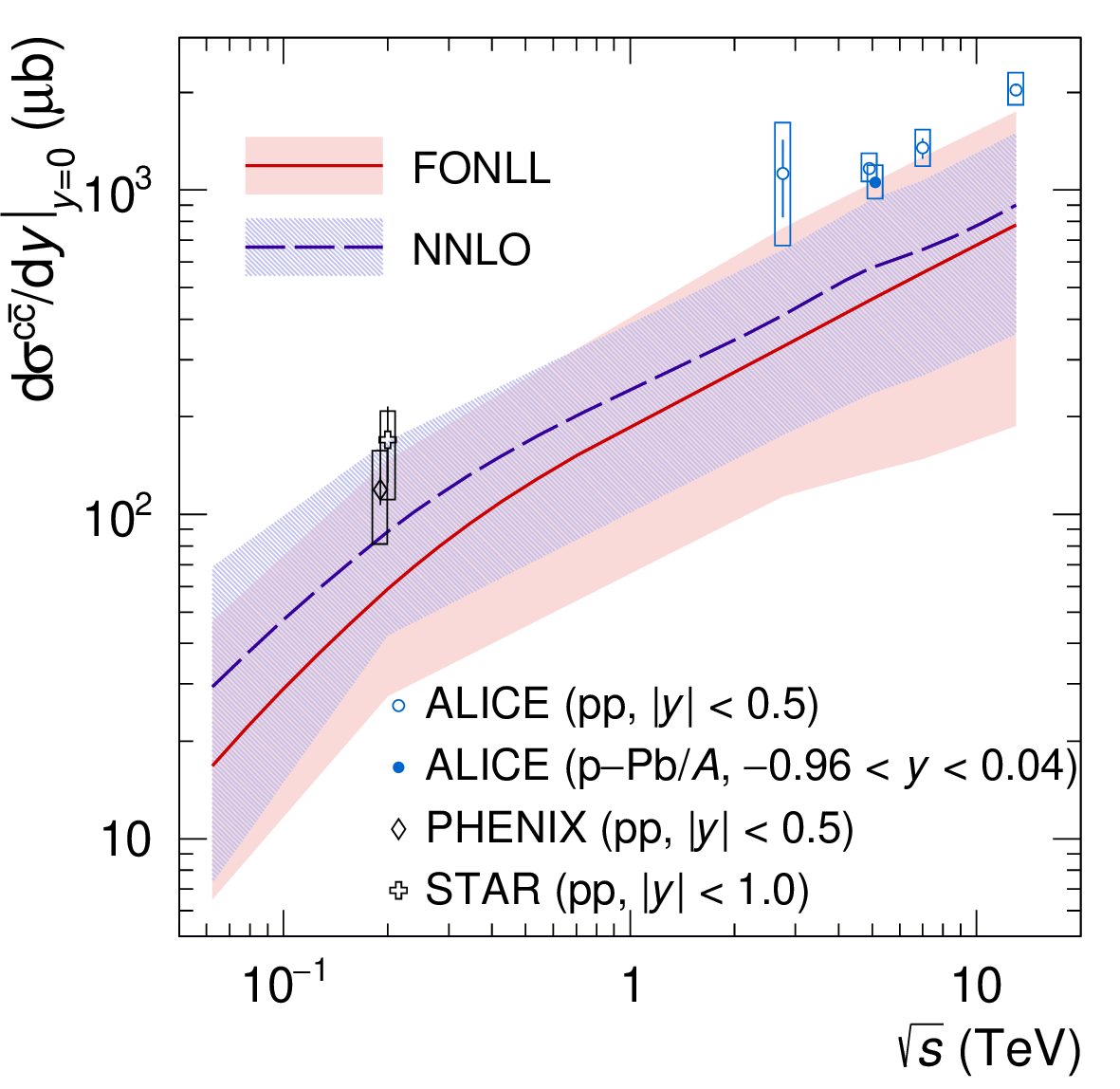}
    \caption{Left: The fragmentation fractions for charm hadrons in \pPb collisions at \sqrtsNNfive, compared with results from pp collisions at \sqrtsfive~\cite{ALICE:2021dhb,ALICE:charmfrag13} and from \ee and \ep collisions at lower energies~\cite{Lisovyi:2015uqa}. The fragmentation fractions for \Jpsi are multiplied by 20 for visibility. Right: Total \ccbar production cross section per unit of rapidity and per nucleon in \pPb collisions, compared with measurements in pp collisions from ALICE~\cite{ALICE:2021dhb,ALICE:charmfrag13}, PHENIX~\cite{PHENIX:2010xji}, and STAR~\cite{STAR:2012nbd}, and with FONLL~\cite{fonllcalc1,fonllcalc2} and NNLO~\cite{nnlo1,nnlo2, nnlo3} pQCD calculations. The statistical uncertainties are shown as bars, and the systematic uncertainties as unfilled boxes.}
    \label{fig:FFandCS}
\end{figure}

In the denominator, only ground-state charm hadrons were considered. The production cross section of \Xicz was considered twice to represent both of the \Xic charge states.The fragmentation fractions of all measured charm-hadron species in \pPb collisions are reported in \figref{fig:FFandCS} (left panel) and summarised in~\tabref{tab:fragfractions}. The results are compared with those measured previously in pp collisions at \sqrtsfive by the ALICE Collaboration. The original measurements in pp collisions published in Ref.~\cite{ALICE:2021dhb} were updated in Ref.~\cite{ALICE:charmfrag13} by taking into account more recent measurements of the \Lc-baryon cross section down to $\pt=0$ and of the prompt \Jpsi cross section. No significant modification of the hadronisation between the two colliding systems is observed despite the larger system size and higher charged-particle multiplicity density in \pPb collisions. 
The values are consistent with those measured in pp collisions at \sqrtsthirt~\cite{ALICE:charmfrag13}.
The measurements are also compared with those from \epem and \ep collisions~\cite{Lisovyi:2015uqa}. As observed for pp collisions, the fragmentation fractions in \pPb collisions also exhibit an overall enhancement in the relative production of charm baryons and a corresponding deficit in charm-meson production with respect to \epem and \ep collisions. 

{\def\arraystretch{1.2}
\begin{table}[h!tb]
    \centering
    \caption{Fragmentation fractions \fragfraction of charm hadrons in pp~\cite{ALICE:charmfrag13} and \pPb collisions at \sqrtsNNfive.}
    \begin{tabular}{c l l }
       \fragfraction (\%)  & pp, \sqrtsfive~\cite{ALICE:charmfrag13} & \pPb, \sqrtsNNfive \\\hline\hline
       \Dz & \ValUncStatSystUD{39.6}{1.7}{2.6}{3.8}& \ValUncStatSystUD{41.3}{1.3}{2.7}{3.2}   \\
       \Dp & \ValUncStatSystUD{17.5}{1.8}{1.7}{2.1} & \ValUncStatSystUD{18.2}{1.0}{1.6}{1.6}  \\
       \Ds & \ValUncStatSystUD{7.4}{1.0}{1.9}{1.1}  & \ValUncStatSystUD{9.0}{0.5}{1.6}{1.2}   \\
       \Lc & \ValUncStatSystUD{18.9}{1.3}{1.5}{2.0}& \ValUncStatSystUD{17.1}{1.1}{1.5}{1.8}   \\
       \Xicz & \ValUncStatSystUD{8.1}{1.2}{2.5}{2.5}& \ValUncStatSystUD{7.0}{1.1}{2.5}{1.8}   \\
       \Jpsi & \ValUncStatSystUD{0.44}{0.03}{0.04}{0.06}& \ValUncStatSystUD{0.41}{0.02}{0.04}{0.04}  \\\hline
       \Dstar  & \ValUncStatSystUD{15.7}{1.2}{4.1}{1.9} &  \ValUncStatSystUD{12.9}{0.6}{1.1}{1.2} 
  
    \end{tabular}
    \label{tab:fragfractions}
\end{table}
}

The total \ccbar production cross section at midrapidity was computed as the sum of the measured \pT-integrated cross sections of the ground-state charm hadrons: \Dz, \Dp, \Ds, \Lc, \Xicz, and \Jpsi. The \Xicz contribution was considered twice to account for the \Xicp cross section, as discussed above. The rapidity shift between the measured charm hadrons and the originating charm-quark pairs was accounted for by multiplying the sum of the hadron species by the aforementioned correction factor of $1.03$. The correlations of the systematic uncertainties between charm-hadron species were treated in the same way as for the fragmentation fractions discussed above. The resulting charm cross section is 

\begin{minipage}{\textwidth}
\begin{align}
\begin{split}
\label{eq:totalxsec}
\left.\frac{\mathrm{d}\sigma(\ccbar)}{\mathrm{d}y}\right|^{\text{p--Pb},\sqrtsNNfive}_{-0.96<y<0.04} = 219.6 \pm 6.3\;(\mathrm{stat.}) {}_{-11.8}^{+10.5}\;(\mathrm{syst.}) {}_{-2.9}^{+8.3}\;(\mathrm{extr.}) \pm 5.4\;(\mathrm{BR})\\[-2ex] {}\pm 4.6\;(\mathrm{lumi.}) \pm 19.5\;(\text{rapidity shape})+15.0\;(\Omegac)~\mb.
\end{split}
\end{align}
\end{minipage}

The measured cross section is compared to results in pp collisions at other centre-of-mass energies from ALICE~\cite{ALICE:2021dhb,ALICE:charmfrag13}, PHENIX~\cite{PHENIX:2010xji}, and STAR~\cite{STAR:2012nbd} in~\figref{fig:FFandCS} (right). In order to be directly comparable with the pp results, the cross section measured in \pPb collisions is scaled down by a factor $1/A$ with $A=208$, under the assumption that the charm production cross section in proton--nucleus collisions scales with the mass number of the larger nucleus due to the binary scaling of hard processes with the number of nucleon--nucleon collisions. The charm production cross section per nucleon in \pPb collisions at \sqrtsNNfive is consistent with the one measured in pp collisions at \sqrtsfive, and is also in agreement with the general trend of results across different collision energies.
The results are compared with pQCD calculations in the FONLL~\cite{fonllcalc1,fonllcalc2} and Next-to-Next-to-Leading Order (NNLO)~\cite{nnlo1,nnlo2,nnlo3} approaches. As observed in previous works relating to the charm production cross section in pp collisions, while the experimental results tend to lie at the upper edge of the theoretical uncertainties, the pQCD approaches describe the evolution of the \ccbar production cross section as a function of \sqrtsNN. 

The \pt-integrated nuclear modification factor \RpPb for \ccbar production in \pPb collisions at \sqrtsNNfive was computed from the total charm cross section discussed above. The \RpPb is calculated from the measured cross sections in \pPb and pp collisions at the same centre-of-mass energy as

\begin{equation}\label{eq:RpPb}
\RpPb=\frac{1}{A}\frac{\left.\mathrm{d}\sigma/\mathrm{d}y\right|^{\mathrm{pPb}}_{-0.96<y<0.04}}{\left.\mathrm{d}\sigma/\mathrm{d}y\right|^{\mathrm{pp}}_{|y|<0.5} \times \alpha_y},
\end{equation}

with $A=208$ as the Pb-nucleus mass number. A value of $\RpPb=1$ would imply that there is no modification of the total yield at midrapidity in \pPb collisions with respect to pp collisions.  The correction factor $\alpha_y=1.01$ accounts for the different $y$ coverage of the measurements in pp and \pPb collisions due to the centre-of-mass rapidity shift between the two collision systems. It was estimated from FONLL calculations of the $\mathrm{d}\sigma/\mathrm{d}y$ distribution of charm quarks. The uncertainties on the measured charm cross sections in pp and \pPb collisions due to the branching ratios and the contribution from \Omegac baryons were treated as fully correlated in the ratio, and the rapidity shape uncertainties as partially correlated. All other sources of uncertainty were propagated as fully uncorrelated.

\begin{figure}[h!tb]
    \centering
    \includegraphics[width=0.65\textwidth]{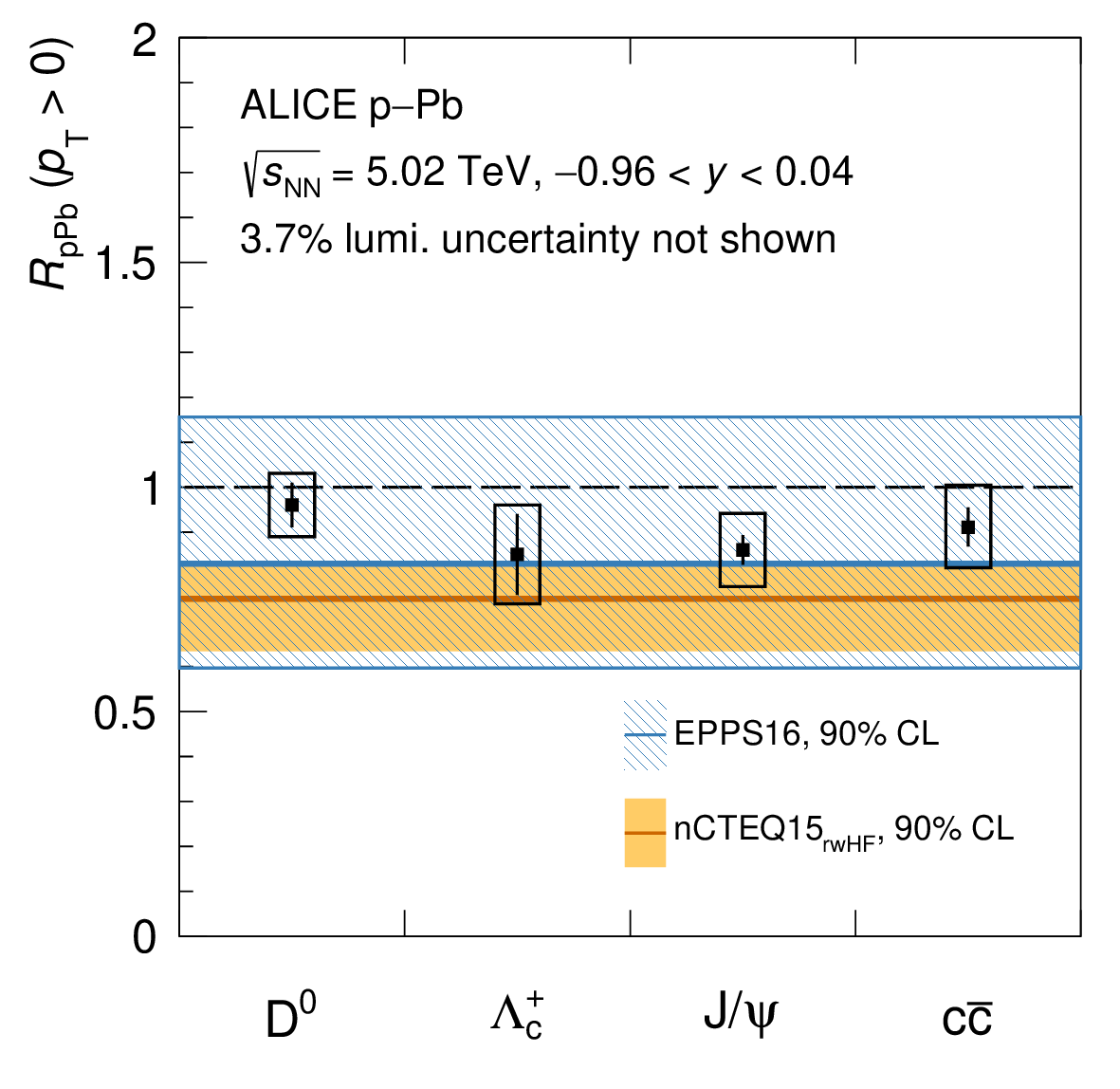}
    \caption{The \pt-integrated nuclear modification factor \RpPb for \Dzero mesons~\cite{ALICE:2019fhe}, \Lc baryons~\cite{ALICE:2022exq}, \Jpsi mesons~\cite{ALICE:2021lmn}, and \ccbar pairs in \pPb collisions at \sqrtsNNfive. The results are compared with calculations using the nCTEQ15$_{\mathrm{rwHF}}$~\cite{Kovarik:2015cma,Kusina:2017gkz} (yellow shaded band) and EPPS16~\cite{Eskola:2016oht} (blue crosshatched band) nPDF sets. The horizontal lines represent the central values of the respective predictions.}
    \label{fig:RpPb}
\end{figure}

The \RpPb for \ccbar pair production at midrapidity at \sqrtsNNfive is shown in \figref{fig:RpPb} and is
\begin{equation}\label{eq:RpPbResult}
\left.\RpPb\right|^{-0.96<y<0.04}_{\pt>0}(\ccbar) = 0.91 \pm 0.04\;{\rm (stat.)}{}^{+0.08}_{-0.09}\;{\rm (syst.)}{}^{+0.05}_{-0.03}\;{\rm (extr.)}{}\pm 0.03\;{\rm (lumi.)},
\end{equation}
which is consistent with unity within 1$\sigma$. This implies that the overall charm production rate in \pPb collisions is consistent with a binary scaling of pp collisions without any significant modification from nuclear effects. 

The result is also compared with those for \Dzero mesons~\cite{ALICE:2019fhe}, \Lc baryons~\cite{ALICE:2022exq}, and \Jpsi mesons~\cite{ALICE:2021lmn}. The \pt-integrated \RpPb values are consistent within uncertainties between all species, implying that the individual particle species' production rates at midrapidity are similarly unaffected by nuclear effects. The results are compared with calculations of the \pt-integrated \Dzero \RpPb using both the EPPS16~\cite{Eskola:2016oht} and nCTEQ15$_{\mathrm{rwHF}}$~\cite{Kovarik:2015cma,Kusina:2017gkz} nPDF sets. In the latter case, the nCTEQ15 predictions are modified by a Bayesian reweighting, constrained to heavy-flavour hadron measurements from the LHC~\cite{Kusina:2017gkz}. The predictions of the \pt-integrated \RpPb are considered to be equivalent for all charm-hadron species, as only the parton distribution functions are varied between the pp and \pPb calculations, and the \pt-integrated fragmentation functions cancel in the ratios. 
The calculation using EPPS16 is consistent with unity within large uncertainties, while the nCTEQ15 nPDFs predict a slight suppression of charm production. The experimental results are described by both of the calculations within uncertainties. In the case of calculations with nCTEQ15, the experimental results are on the upper edge of the model uncertainty. 

\section{Summary} \label{sec:summary}

The total charm production cross section per unit of rapidity and charm-hadron fragmentation fractions were measured for the first time at midrapidity in \pPb collisions at \sqrtsNNfive. The fragmentation fractions for all charm-hadron species were found to be consistent for pp and \pPb collisions at the same collision energy, indicating that there is no significant modification of the charm-quark hadronisation process due to the different hadronic collisions' system sizes. 
The total charm production cross section per nucleon in \pPb collisions was measured to be consistent with pp collisions at the same centre-of-mass energy within the uncertainties, meaning that the assumption of binary scaling holds for charm-quark production. This was further confirmed by examining the nuclear modification factor \RpPb, which has a value consistent with unity at midrapidity and compatible with both the individual \pt-integrated charm-hadron \RpPb results and with model calculations including nuclear modification of the PDFs. 
Future measurements during Run 3 of the LHC will allow for a reduction of the uncertainties possibly reducing the reliance on model-dependent extrapolations.  
A particular focus in this field will be the determination of charm and beauty fragmentation fractions in \PbPb collisions, in order to precisely quantify modification of hadronisation driven by the formation and presence of a quark--gluon plasma, like modification of the abundance of different charm-hadron species~\cite{ALICE:2021kfc, STAR:2021tte,ALICE:2021bib,STAR:2019ank} and quarkonium states~\cite{ALICE:2022jeh,ALICE:2023gco}. Future measurements in \pPb collisions with higher precision will further allow significant constraints to be drawn on nuclear modifications of the PDFs.


\newenvironment{acknowledgement}{\relax}{\relax}
\begin{acknowledgement}
\section*{Acknowledgements}

The ALICE Collaboration would like to thank all its engineers and technicians for their invaluable contributions to the construction of the experiment and the CERN accelerator teams for the outstanding performance of the LHC complex.
The ALICE Collaboration gratefully acknowledges the resources and support provided by all Grid centres and the Worldwide LHC Computing Grid (WLCG) collaboration.
The ALICE Collaboration acknowledges the following funding agencies for their support in building and running the ALICE detector:
A. I. Alikhanyan National Science Laboratory (Yerevan Physics Institute) Foundation (ANSL), State Committee of Science and World Federation of Scientists (WFS), Armenia;
Austrian Academy of Sciences, Austrian Science Fund (FWF): [M 2467-N36] and Nationalstiftung f\"{u}r Forschung, Technologie und Entwicklung, Austria;
Ministry of Communications and High Technologies, National Nuclear Research Center, Azerbaijan;
Conselho Nacional de Desenvolvimento Cient\'{\i}fico e Tecnol\'{o}gico (CNPq), Financiadora de Estudos e Projetos (Finep), Funda\c{c}\~{a}o de Amparo \`{a} Pesquisa do Estado de S\~{a}o Paulo (FAPESP) and Universidade Federal do Rio Grande do Sul (UFRGS), Brazil;
Bulgarian Ministry of Education and Science, within the National Roadmap for Research Infrastructures 2020-2027 (object CERN), Bulgaria;
Ministry of Education of China (MOEC) , Ministry of Science \& Technology of China (MSTC) and National Natural Science Foundation of China (NSFC), China;
Ministry of Science and Education and Croatian Science Foundation, Croatia;
Centro de Aplicaciones Tecnol\'{o}gicas y Desarrollo Nuclear (CEADEN), Cubaenerg\'{\i}a, Cuba;
Ministry of Education, Youth and Sports of the Czech Republic, Czech Republic;
The Danish Council for Independent Research | Natural Sciences, the VILLUM FONDEN and Danish National Research Foundation (DNRF), Denmark;
Helsinki Institute of Physics (HIP), Finland;
Commissariat \`{a} l'Energie Atomique (CEA) and Institut National de Physique Nucl\'{e}aire et de Physique des Particules (IN2P3) and Centre National de la Recherche Scientifique (CNRS), France;
Bundesministerium f\"{u}r Bildung und Forschung (BMBF) and GSI Helmholtzzentrum f\"{u}r Schwerionenforschung GmbH, Germany;
General Secretariat for Research and Technology, Ministry of Education, Research and Religions, Greece;
National Research, Development and Innovation Office, Hungary;
Department of Atomic Energy Government of India (DAE), Department of Science and Technology, Government of India (DST), University Grants Commission, Government of India (UGC) and Council of Scientific and Industrial Research (CSIR), India;
National Research and Innovation Agency - BRIN, Indonesia;
Istituto Nazionale di Fisica Nucleare (INFN), Italy;
Japanese Ministry of Education, Culture, Sports, Science and Technology (MEXT) and Japan Society for the Promotion of Science (JSPS) KAKENHI, Japan;
Consejo Nacional de Ciencia (CONACYT) y Tecnolog\'{i}a, through Fondo de Cooperaci\'{o}n Internacional en Ciencia y Tecnolog\'{i}a (FONCICYT) and Direcci\'{o}n General de Asuntos del Personal Academico (DGAPA), Mexico;
Nederlandse Organisatie voor Wetenschappelijk Onderzoek (NWO), Netherlands;
The Research Council of Norway, Norway;
Pontificia Universidad Cat\'{o}lica del Per\'{u}, Peru;
Ministry of Science and Higher Education, National Science Centre and WUT ID-UB, Poland;
Korea Institute of Science and Technology Information and National Research Foundation of Korea (NRF), Republic of Korea;
Ministry of Education and Scientific Research, Institute of Atomic Physics, Ministry of Research and Innovation and Institute of Atomic Physics and Universitatea Nationala de Stiinta si Tehnologie Politehnica Bucuresti, Romania;
Ministry of Education, Science, Research and Sport of the Slovak Republic, Slovakia;
National Research Foundation of South Africa, South Africa;
Swedish Research Council (VR) and Knut \& Alice Wallenberg Foundation (KAW), Sweden;
European Organization for Nuclear Research, Switzerland;
Suranaree University of Technology (SUT), National Science and Technology Development Agency (NSTDA) and National Science, Research and Innovation Fund (NSRF via PMU-B B05F650021), Thailand;
Turkish Energy, Nuclear and Mineral Research Agency (TENMAK), Turkey;
National Academy of  Sciences of Ukraine, Ukraine;
Science and Technology Facilities Council (STFC), United Kingdom;
National Science Foundation of the United States of America (NSF) and United States Department of Energy, Office of Nuclear Physics (DOE NP), United States of America.
In addition, individual groups or members have received support from:
Czech Science Foundation (grant no. 23-07499S), Czech Republic;
European Research Council (grant no. 950692), European Union;
ICSC - Centro Nazionale di Ricerca in High Performance Computing, Big Data and Quantum Computing, European Union - NextGenerationEU;
Academy of Finland (Center of Excellence in Quark Matter) (grant nos. 346327, 346328), Finland.

\end{acknowledgement}

\bibliographystyle{utphys}   
\bibliography{bibliography}

\newpage
\appendix

%
%

\section{The ALICE Collaboration}
\label{app:collab}
\begin{flushleft} 
\small

S.~Acharya\,\orcidlink{0000-0002-9213-5329}\,$^{\rm 127}$, 
D.~Adamov\'{a}\,\orcidlink{0000-0002-0504-7428}\,$^{\rm 86}$, 
A.~Agarwal$^{\rm 135}$, 
G.~Aglieri Rinella\,\orcidlink{0000-0002-9611-3696}\,$^{\rm 32}$, 
L.~Aglietta\,\orcidlink{0009-0003-0763-6802}\,$^{\rm 24}$, 
M.~Agnello\,\orcidlink{0000-0002-0760-5075}\,$^{\rm 29}$, 
N.~Agrawal\,\orcidlink{0000-0003-0348-9836}\,$^{\rm 25}$, 
Z.~Ahammed\,\orcidlink{0000-0001-5241-7412}\,$^{\rm 135}$, 
S.~Ahmad\,\orcidlink{0000-0003-0497-5705}\,$^{\rm 15}$, 
S.U.~Ahn\,\orcidlink{0000-0001-8847-489X}\,$^{\rm 71}$
% ...

\section*{Affiliation Notes}

$^{\rm I}$ Deceased\\
$^{\rm II}$ Also at: Max-Planck-Institut fur Physik, Munich, Germany\\
$^{\rm III}$ Also at: Italian National Agency for New Technologies, Energy and Sustainable Economic Development (ENEA), Bologna, Italy\\
$^{\rm IV}$ Also at: Dipartimento DET del Politecnico di Torino, Turin, Italy\\
$^{\rm V}$ Also at: Yildiz Technical University, Istanbul, T\"{u}rkiye\\
$^{\rm VI}$ Also at: Department of Applied Physics, Aligarh Muslim University, Aligarh, India\\
$^{\rm VII}$ Also at: Institute of Theoretical Physics, University of Wroclaw, Poland\\
$^{\rm VIII}$ Also at: An institution covered by a cooperation agreement with CERN\\

\section*{Collaboration Institutes}

$^{1}$ A.I. Alikhanyan National Science Laboratory (Yerevan Physics Institute) Foundation, Yerevan, Armenia\\
$^{2}$ AGH University of Krakow, Cracow, Poland\\
$^{3}$ Bogolyubov Institute for Theoretical Physics, National Academy of Sciences of Ukraine, Kiev, Ukraine\\
$^{4}$ Bose Institute, Department of Physics  and Centre for Astroparticle Physics and Space Science (CAPSS), Kolkata, India\\
$^{5}$ California Polytechnic State University, San Luis Obispo, California, United States\\
$^{6}$ Central China Normal University, Wuhan, China\\
$^{7}$ Centro de Aplicaciones Tecnol\'{o}gicas y Desarrollo Nuclear (CEADEN), Havana, Cuba\\
$^{8}$ Centro de Investigaci\'{o}n y de Estudios Avanzados (CINVESTAV), Mexico City and M\'{e}rida, Mexico\\
$^{9}$ Chicago State University, Chicago, Illinois, United States\\
$^{10}$ China Institute of Atomic Energy, Beijing, China\\
$^{11}$ China University of Geosciences, Wuhan, China\\
$^{12}$ Chungbuk National University, Cheongju, Republic of Korea\\
$^{13}$ Comenius University Bratislava, Faculty of Mathematics, Physics and Informatics, Bratislava, Slovak Republic\\
$^{14}$ Creighton University, Omaha, Nebraska, United States\\
$^{15}$ Department of Physics, Aligarh Muslim University, Aligarh, India\\
$^{16}$ Department of Physics, Pusan National University, Pusan, Republic of Korea\\
$^{17}$ Department of Physics, Sejong University, Seoul, Republic of Korea\\
$^{18}$ Department of Physics, University of California, Berkeley, California, United States\\
$^{19}$ Department of Physics, University of Oslo, Oslo, Norway\\
$^{20}$ Department of Physics and Technology, University of Bergen, Bergen, Norway\\
$^{21}$ Dipartimento di Fisica, Universit\`{a} di Pavia, Pavia, Italy\\
$^{22}$ Dipartimento di Fisica dell'Universit\`{a} and Sezione INFN, Cagliari, Italy\\
$^{23}$ Dipartimento di Fisica dell'Universit\`{a} and Sezione INFN, Trieste, Italy\\
$^{24}$ Dipartimento di Fisica dell'Universit\`{a} and Sezione INFN, Turin, Italy\\
$^{25}$ Dipartimento di Fisica e Astronomia dell'Universit\`{a} and Sezione INFN, Bologna, Italy\\
$^{26}$ Dipartimento di Fisica e Astronomia dell'Universit\`{a} and Sezione INFN, Catania, Italy\\
$^{27}$ Dipartimento di Fisica e Astronomia dell'Universit\`{a} and Sezione INFN, Padova, Italy\\
$^{28}$ Dipartimento di Fisica `E.R.~Caianiello' dell'Universit\`{a} and Gruppo Collegato INFN, Salerno, Italy\\
$^{29}$ Dipartimento DISAT del Politecnico and Sezione INFN, Turin, Italy\\
$^{30}$ Dipartimento di Scienze MIFT, Universit\`{a} di Messina, Messina, Italy\\
$^{31}$ Dipartimento Interateneo di Fisica `M.~Merlin' and Sezione INFN, Bari, Italy\\
$^{32}$ European Organization for Nuclear Research (CERN), Geneva, Switzerland\\
$^{33}$ Faculty of Electrical Engineering, Mechanical Engineering and Naval Architecture, University of Split, Split, Croatia\\
$^{34}$ Faculty of Engineering and Science, Western Norway University of Applied Sciences, Bergen, Norway\\
$^{35}$ Faculty of Nuclear Sciences and Physical Engineering, Czech Technical University in Prague, Prague, Czech Republic\\
$^{36}$ Faculty of Physics, Sofia University, Sofia, Bulgaria\\
$^{37}$ Faculty of Science, P.J.~\v{S}af\'{a}rik University, Ko\v{s}ice, Slovak Republic\\
$^{38}$ Frankfurt Institute for Advanced Studies, Johann Wolfgang Goethe-Universit\"{a}t Frankfurt, Frankfurt, Germany\\
$^{39}$ Fudan University, Shanghai, China\\
$^{40}$ Gangneung-Wonju National University, Gangneung, Republic of Korea\\
$^{41}$ Gauhati University, Department of Physics, Guwahati, India\\
$^{42}$ Helmholtz-Institut f\"{u}r Strahlen- und Kernphysik, Rheinische Friedrich-Wilhelms-Universit\"{a}t Bonn, Bonn, Germany\\
$^{43}$ Helsinki Institute of Physics (HIP), Helsinki, Finland\\
$^{44}$ High Energy Physics Group,  Universidad Aut\'{o}noma de Puebla, Puebla, Mexico\\
$^{45}$ Horia Hulubei National Institute of Physics and Nuclear Engineering, Bucharest, Romania\\
$^{46}$ HUN-REN Wigner Research Centre for Physics, Budapest, Hungary\\
$^{47}$ Indian Institute of Technology Bombay (IIT), Mumbai, India\\
$^{48}$ Indian Institute of Technology Indore, Indore, India\\
$^{49}$ INFN, Laboratori Nazionali di Frascati, Frascati, Italy\\
$^{50}$ INFN, Sezione di Bari, Bari, Italy\\
$^{51}$ INFN, Sezione di Bologna, Bologna, Italy\\
$^{52}$ INFN, Sezione di Cagliari, Cagliari, Italy\\
$^{53}$ INFN, Sezione di Catania, Catania, Italy\\
$^{54}$ INFN, Sezione di Padova, Padova, Italy\\
$^{55}$ INFN, Sezione di Pavia, Pavia, Italy\\
$^{56}$ INFN, Sezione di Torino, Turin, Italy\\
$^{57}$ INFN, Sezione di Trieste, Trieste, Italy\\
$^{58}$ Inha University, Incheon, Republic of Korea\\
$^{59}$ Institute for Gravitational and Subatomic Physics (GRASP), Utrecht University/Nikhef, Utrecht, Netherlands\\
$^{60}$ Institute of Experimental Physics, Slovak Academy of Sciences, Ko\v{s}ice, Slovak Republic\\
$^{61}$ Institute of Physics, Homi Bhabha National Institute, Bhubaneswar, India\\
$^{62}$ Institute of Physics of the Czech Academy of Sciences, Prague, Czech Republic\\
$^{63}$ Institute of Space Science (ISS), Bucharest, Romania\\
$^{64}$ Institut f\"{u}r Kernphysik, Johann Wolfgang Goethe-Universit\"{a}t Frankfurt, Frankfurt, Germany\\
$^{65}$ Instituto de Ciencias Nucleares, Universidad Nacional Aut\'{o}noma de M\'{e}xico, Mexico City, Mexico\\
$^{66}$ Instituto de F\'{i}sica, Universidade Federal do Rio Grande do Sul (UFRGS), Porto Alegre, Brazil\\
$^{67}$ Instituto de F\'{\i}sica, Universidad Nacional Aut\'{o}noma de M\'{e}xico, Mexico City, Mexico\\
$^{68}$ iThemba LABS, National Research Foundation, Somerset West, South Africa\\
$^{69}$ Jeonbuk National University, Jeonju, Republic of Korea\\
$^{70}$ Johann-Wolfgang-Goethe Universit\"{a}t Frankfurt Institut f\"{u}r Informatik, Fachbereich Informatik und Mathematik, Frankfurt, Germany\\
$^{71}$ Korea Institute of Science and Technology Information, Daejeon, Republic of Korea\\
$^{72}$ KTO Karatay University, Konya, Turkey\\
$^{73}$ Laboratoire de Physique Subatomique et de Cosmologie, Universit\'{e} Grenoble-Alpes, CNRS-IN2P3, Grenoble, France\\
$^{74}$ Lawrence Berkeley National Laboratory, Berkeley, California, United States\\
$^{75}$ Lund University Department of Physics, Division of Particle Physics, Lund, Sweden\\
$^{76}$ Nagasaki Institute of Applied Science, Nagasaki, Japan\\
$^{77}$ Nara Women{'}s University (NWU), Nara, Japan\\
$^{78}$ National and Kapodistrian University of Athens, School of Science, Department of Physics , Athens, Greece\\
$^{79}$ National Centre for Nuclear Research, Warsaw, Poland\\
$^{80}$ National Institute of Science Education and Research, Homi Bhabha National Institute, Jatni, India\\
$^{81}$ National Nuclear Research Center, Baku, Azerbaijan\\
$^{82}$ National Research and Innovation Agency - BRIN, Jakarta, Indonesia\\
$^{83}$ Niels Bohr Institute, University of Copenhagen, Copenhagen, Denmark\\
$^{84}$ Nikhef, National institute for subatomic physics, Amsterdam, Netherlands\\
$^{85}$ Nuclear Physics Group, STFC Daresbury Laboratory, Daresbury, United Kingdom\\
$^{86}$ Nuclear Physics Institute of the Czech Academy of Sciences, Husinec-\v{R}e\v{z}, Czech Republic\\
$^{87}$ Oak Ridge National Laboratory, Oak Ridge, Tennessee, United States\\
$^{88}$ Ohio State University, Columbus, Ohio, United States\\
$^{89}$ Physics department, Faculty of science, University of Zagreb, Zagreb, Croatia\\
$^{90}$ Physics Department, Panjab University, Chandigarh, India\\
$^{91}$ Physics Department, University of Jammu, Jammu, India\\
$^{92}$ Physics Program and International Institute for Sustainability with Knotted Chiral Meta Matter (SKCM2), Hiroshima University, Hiroshima, Japan\\
$^{93}$ Physikalisches Institut, Eberhard-Karls-Universit\"{a}t T\"{u}bingen, T\"{u}bingen, Germany\\
$^{94}$ Physikalisches Institut, Ruprecht-Karls-Universit\"{a}t Heidelberg, Heidelberg, Germany\\
$^{95}$ Physik Department, Technische Universit\"{a}t M\"{u}nchen, Munich, Germany\\
$^{96}$ Politecnico di Bari and Sezione INFN, Bari, Italy\\
$^{97}$ Research Division and ExtreMe Matter Institute EMMI, GSI Helmholtzzentrum f\"ur Schwerionenforschung GmbH, Darmstadt, Germany\\
$^{98}$ Saga University, Saga, Japan\\
$^{99}$ Saha Institute of Nuclear Physics, Homi Bhabha National Institute, Kolkata, India\\
$^{100}$ School of Physics and Astronomy, University of Birmingham, Birmingham, United Kingdom\\
$^{101}$ Secci\'{o}n F\'{\i}sica, Departamento de Ciencias, Pontificia Universidad Cat\'{o}lica del Per\'{u}, Lima, Peru\\
$^{102}$ Stefan Meyer Institut f\"{u}r Subatomare Physik (SMI), Vienna, Austria\\
$^{103}$ SUBATECH, IMT Atlantique, Nantes Universit\'{e}, CNRS-IN2P3, Nantes, France\\
$^{104}$ Sungkyunkwan University, Suwon City, Republic of Korea\\
$^{105}$ Suranaree University of Technology, Nakhon Ratchasima, Thailand\\
$^{106}$ Technical University of Ko\v{s}ice, Ko\v{s}ice, Slovak Republic\\
$^{107}$ The Henryk Niewodniczanski Institute of Nuclear Physics, Polish Academy of Sciences, Cracow, Poland\\
$^{108}$ The University of Texas at Austin, Austin, Texas, United States\\
$^{109}$ Universidad Aut\'{o}noma de Sinaloa, Culiac\'{a}n, Mexico\\
$^{110}$ Universidade de S\~{a}o Paulo (USP), S\~{a}o Paulo, Brazil\\
$^{111}$ Universidade Estadual de Campinas (UNICAMP), Campinas, Brazil\\
$^{112}$ Universidade Federal do ABC, Santo Andre, Brazil\\
$^{113}$ Universitatea Nationala de Stiinta si Tehnologie Politehnica Bucuresti, Bucharest, Romania\\
$^{114}$ University of Cape Town, Cape Town, South Africa\\
$^{115}$ University of Derby, Derby, United Kingdom\\
$^{116}$ University of Houston, Houston, Texas, United States\\
$^{117}$ University of Jyv\"{a}skyl\"{a}, Jyv\"{a}skyl\"{a}, Finland\\
$^{118}$ University of Kansas, Lawrence, Kansas, United States\\
$^{119}$ University of Liverpool, Liverpool, United Kingdom\\
$^{120}$ University of Science and Technology of China, Hefei, China\\
$^{121}$ University of South-Eastern Norway, Kongsberg, Norway\\
$^{122}$ University of Tennessee, Knoxville, Tennessee, United States\\
$^{123}$ University of the Witwatersrand, Johannesburg, South Africa\\
$^{124}$ University of Tokyo, Tokyo, Japan\\
$^{125}$ University of Tsukuba, Tsukuba, Japan\\
$^{126}$ Universit\"{a}t M\"{u}nster, Institut f\"{u}r Kernphysik, M\"{u}nster, Germany\\
$^{127}$ Universit\'{e} Clermont Auvergne, CNRS/IN2P3, LPC, Clermont-Ferrand, France\\
$^{128}$ Universit\'{e} de Lyon, CNRS/IN2P3, Institut de Physique des 2 Infinis de Lyon, Lyon, France\\
$^{129}$ Universit\'{e} de Strasbourg, CNRS, IPHC UMR 7178, F-67000 Strasbourg, France, Strasbourg, France\\
$^{130}$ Universit\'{e} Paris-Saclay, Centre d'Etudes de Saclay (CEA), IRFU, D\'{e}partment de Physique Nucl\'{e}aire (DPhN), Saclay, France\\
$^{131}$ Universit\'{e}  Paris-Saclay, CNRS/IN2P3, IJCLab, Orsay, France\\
$^{132}$ Universit\`{a} degli Studi di Foggia, Foggia, Italy\\
$^{133}$ Universit\`{a} del Piemonte Orientale, Vercelli, Italy\\
$^{134}$ Universit\`{a} di Brescia, Brescia, Italy\\
$^{135}$ Variable Energy Cyclotron Centre, Homi Bhabha National Institute, Kolkata, India\\
$^{136}$ Warsaw University of Technology, Warsaw, Poland\\
$^{137}$ Wayne State University, Detroit, Michigan, United States\\
$^{138}$ Yale University, New Haven, Connecticut, United States\\
$^{139}$ Yonsei University, Seoul, Republic of Korea\\
$^{140}$  Zentrum  f\"{u}r Technologie und Transfer (ZTT), Worms, Germany\\
$^{141}$ Affiliated with an institute covered by a cooperation agreement with CERN\\
$^{142}$ Affiliated with an international laboratory covered by a cooperation agreement with CERN.\\

\end{flushleft} 

\end{document}